\documentclass{emulateapj_rj}

\shorttitle{DISTANCE TO NGC~4258}
\shortauthors{MAGER, MADORE \& FREEDMAN}

\begin{document}

\title{METALLICITY-CORRECTED TIP OF THE RED GIANT BRANCH DISTANCE TO 
NGC~4258}

\author{\bf Violet A. Mager,  Barry F. Madore \&  Wendy L. Freedman}
\affil{The Observatories\\ Carnegie Institution of Washington\\ 813 Santa Barbara St.\\
    Pasadena, CA 91101}
\email{vmager@ociw.edu, barry@ociw.edu, wendy@ociw.edu}

\begin{abstract}
We have determined the distance to NGC~4258 using observations made
with the Hubble Space Telescope (HST) and the Wide Field, Advanced
Camera for Surveys (ACS/WFC). We apply a modified technique that fully
accounts for metallicity effects on the use of the luminosity of the
tip of the red giant branch (TRGB) to determine one of the most
precise TRGB distance moduli to date:
$\mu(TRGB) = 29.28 \pm 0.04$ (random) $\pm 0.12$ (systematic) mag
($7.18 \pm 0.13 \pm 0.40$ Mpc). We
discuss this distance modulus with respect to other recent
applications of the TRGB method to NGC~4258, and with several other
techniques (Cepheids and masers) that are equally competitive in their
precision, but different in their systematics. 
\end{abstract}


\section{Introduction}

This is the first in a short series of papers using a refined
methodology for determining distances using the discontinuity in the 
I-band magnitude of the red giant branch luminosity function as a standard
candle (Lee et al. 1993), the so-called TRGB (tip of the red giant branch) 
method. Here we apply a
new methodology in correcting for the now well understood and
precisely calibrated metallicity effects on the TRGB magnitude
(see Section 4.1 and Madore et al. 2008).

Our first target is the spiral galaxy NGC~4258.
It is nearby, and therefore very highly resolved, not only into
its bright, high-mass Population I disk stars, but also into its
fainter, but still accessible, low-mass Population II halo
stars. NGC~4258 contains many known Cepheids that have been discovered
and used as distance indicators in multiple observing campaigns using
HST. Its halo has been resolved and studied on equally as many
occasions, revealing a broad, richly populated giant branch for
TRGB distance determination. The uniqueness of NGC~4258
lies at its center, where a Keplerian-rotating
disk of water masers has proper motions and radial
velocities that can be cross-compared and modeled with essentially one
additional free parameter: the distance. As such, the independently
calibrated Population~I (Cepheid) and Population~II (TRGB) distance scales 
both converge on and cross at NGC~4258, where they can be compared to that
from simple geometry (maser method). No other galaxy provides such an 
environment for testing
the distance scale. That said, it must also be emphasized that
NGC~4258 is still only one object, and its uniqueness means that there
is no independent check on the maser distance methodology itself, its
random errors, or its systematics.

Without prejudice as to which (if any) of the three distance
determination methods discussed here is better (understood or
calibrated) at this point, we now proceed to present a new and improved
determination of the TRGB distance using HST ACS/WFC data from one of
our approved and scheduled programs, and archival WFPC2 data as a
consistency check. We compare these results with previous TRGB results, 
and with the other past and published methods.

\section{Data Reduction and Calibration of the ACS Images}

Our HST/ACS observations of the NGC~4258 halo (PID 9477, PI Madore, B.F.)
consist of $2\times 2850$~s exposures in F555W, and $2\times 1300$~s
exposures in F814W. 
Figure 1 shows the location of the fields-of-view of the ACS 
(thick-lined polygon) and partially over-lapping archival WFPC2 (thin-lined
polygons) observations, overlayed on a 
DSS\footnote{The Digitized Sky Surveys were
produced at the Space Telescope Science Institute under
U.S. Government grant NAG W-2166.  The images of these surveys are
based on photographic data obtained using the Oschin Schmidt Telescope
on Palomar Mountain and the UK Schmidt Telescope. The plates were
processed into the present compressed digital form with the permission
of these institutions.} image of NGC 4258. As a consistency check, we used 
two different methods of identifying the stars and calibrating the 
photometry. Both produced TRGB magnitudes that agree to 
within the uncertainties. The details of each method are described below. 


\epsscale{1.00}
\noindent
\begin{figure*}
\plotone{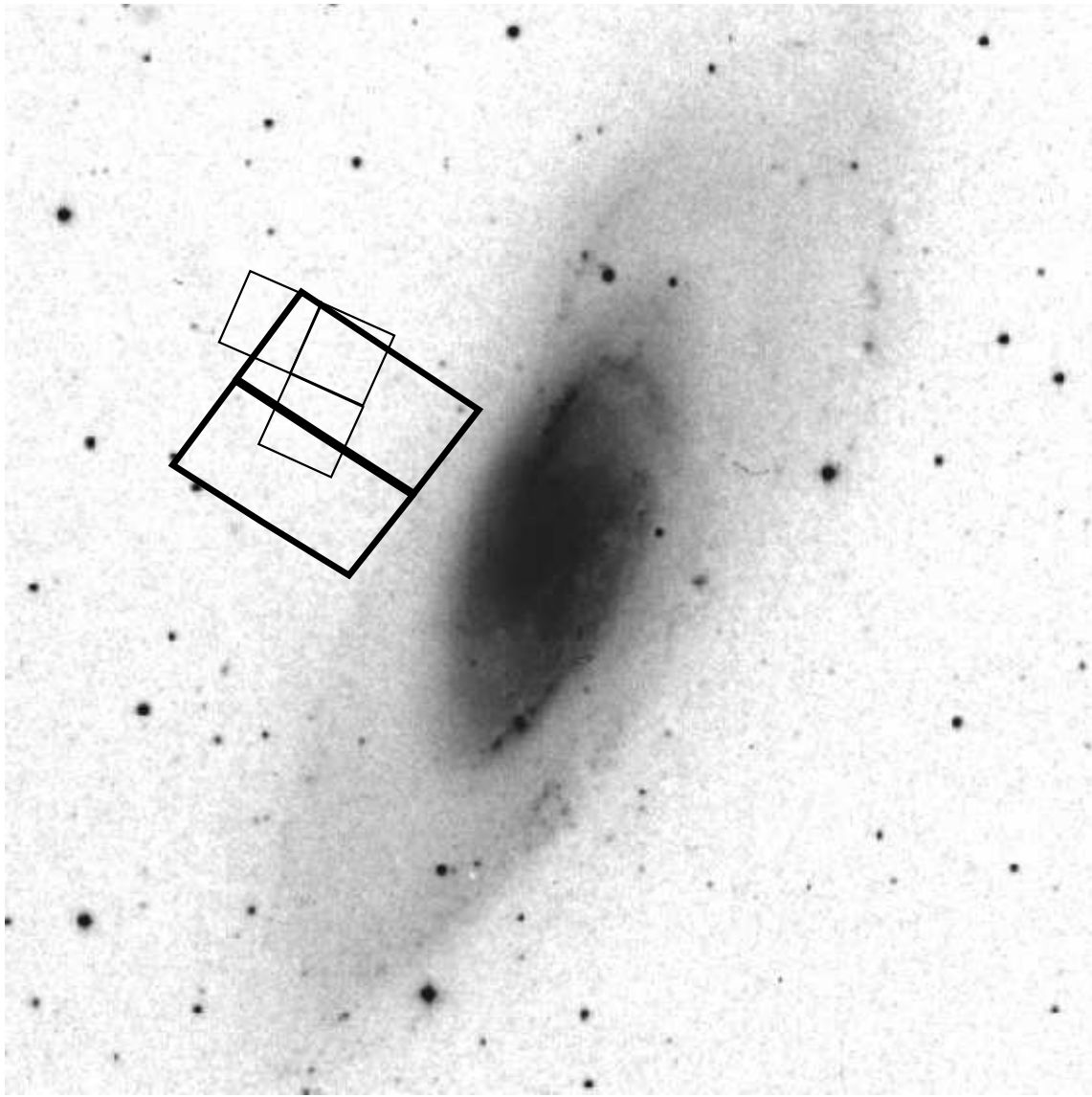}
\caption{DSS image of NGC~4258 overlayed with the footprints of
the HST ACS (thick-lined polygon) and WFPC2 (thin-lined polygons) images
used in this analysis. North is up, and east is to the left.}
\end{figure*}


For the first method, we identified stars and calibrated the ACS images 
using the ACS module
for the highly automated DOLPHOT package\footnote{By Andrew Dolphin,
\url{http://purcell.as.arizona.edu/dolphot/}} (see the DOLPHOT User's
Guide for
details.\footnote{\url{http://purcell.as.arizona.edu/dolphot/dolphot.ps.gz}})
We applied the DOLPHOT package to the STSCI pipeline-processed and
cosmic-ray cleaned images of NGC~4258, masking bad pixels and using
the recommended settings given in the DOLPHOT/ACS User's
Guide\footnote{\url{http://purcell.as.arizona.edu/dolphot/dolphotACS.ps.gz}}.
In order to additionally reject non-stellar objects and objects with
highly uncertain photometry, we selected only those detections with a
DOLPHOT output type of 1 (determined by DOLPHOT to be a "good star"),
with a flag of 0 (the star was "recovered well" in the image), with
sharpness measurements of --0.3 to +0.3, and with a crowding parameter
$< 0.5$ mag.  We then applied Galactic extinction corrections using
the E(B--V) reddening measurements given in the NASA/IPAC
Extragalactic Database (NED), and the total-to-selective absorption
ratios (A/E(B--V)) presented in Sirianni et al. (2005).

As a consistency check on the photometry, we also used DAOPHOT II
(Stetson 1987) and ALLSTAR (Stetson 1994) to independently reduce 
and analyze the ACS images. We located stars within $5\sigma$ of the 
sky in the F814W images, applying sharpness cut-offs determined through 
visual inspection of the results to reject bad pixels and low surface 
brightness galaxies, as well as roundness cut-offs to reject bad rows, 
columns, and highly inclined galaxies. Any remaining bad pixels were 
rejected when objects whose PSF-fit photometry could not be determined 
for both the F814W and the F555W images were eliminated from our analysis.
We rejected most of the remaining non-stellar objects (cosmic rays
and/or compact galaxies that coincidentally had PSF's similar to the
stellar PSF) by applying a cut-off limit in the ALLSTAR $\chi$
measurement (which gives an indication of how good the PSF fit was
with respect to the other objects in the group it was measured
in). After visual inspection of its effects on the color-magnitude
diagram (CMD), we chose $\chi_V < 2$ and $\chi_I < 3$.
Aperture corrections and transformations of the F555W and F814W magnitudes 
to Johnson V and I, respectively, were applied following the method 
outlined in Sirianni et al. (2005). We then
applied the Galactic extinction corrections as described above.

Method 1 (DOLPHOT) and Method 2 (DAOPHOT II/ALLSTAR)) both produced
consistent photometry at the brightness level of the TRGB. Our
edge-detection software (as described in Section 4.2) found that the
I-band magnitude of the TRGB for both photometry methods agrees to
within 0.04 mag, which is within the measured uncertainties. However,
we note in passing that DOLPHOT went deeper than the $5\sigma$ cut-off
used in Method 2, and reported somewhat smaller photometric
uncertainties at the tip. As such, we present the photometry from
DOLPHOT in all subsequent figures and analysis of the ACS data.

\section{Data Reduction and Calibration of the WFPC2 Images}

As a comparison, we downloaded pipeline-calibrated WFPC2 images of the 
NGC~4258 halo (PID 9086, PI Ferguson, H.C.) from the STScI (Space 
Telescope Science Institute) archive (see Fig. 1 for the location of these
observations). These observations include $9\times 1268$~s (on
average) exposures in F606W, and $9\times 1300$~s exposures in F814W.
We did not use the smaller field-of-view PC chip in our analysis. We
registered the individual images of each of the chips in each filter
with the \texttt{IMSHIFT} task in IRAF\footnote{IRAF is distributed by
the National Optical Astronomy Observatories, which are operated by
the Association of Universities for Research in Astronomy, Inc., under
cooperative agreement with the National Science Foundation.}, and
averaged the images (with cosmic ray rejection) using
IRAF/\texttt{IMCOMBINE}. As with Method 2 for the ACS data (Section
2), we used DAOPHOT II and ALLSTAR to locate stars above $2\sigma$
of the sky, and to determine their PSF-fit magnitudes.

We applied transformations and corrections as follows. The
instrumental F606W and F814W magnitudes were CTE (charge transfer
efficiency)-loss corrected, aperture corrected, and transformed to 
Johnson V and I magnitudes (respectively) using the methods outlined 
in Dolphin (2000), with the updated 2002 calibration constants.\footnote{
http://purcell.as.arizona.edu/wfpc2\_calib/} The WFPC2
first- and second-order color terms are much larger
than those for ACS, and required five iterations of the
transformation equation to converge at the 0.001~mag level for
each star. We further rejected non-stellar detections by requiring
that $\chi_V < 0.9$ and $\chi_I < 1$ (see Section 2). Finally,
we corrected for Galactic extinction using the
E(B--V) = 0.016~mag reddening value calculated by NED for the Galactic
line of sight to NGC~4258, and transformed to  A$_V$ = $2.68\times$
E(B--V) and A$_I$ = $1.82\times$ E(B--V).

\section{Measuring the Apparent Magnitude of the TRGB}

Figure 2 shows the CMD's for NGC~4258 from the WFPC2 data (left panel)
and the ACS data (right panel). The error bar in the lower right
corner of each panel represents the median uncertainty on the
photometry of the stars within $\pm 0.3$ mag of the location of the
TRGB. The comparison of data sets is impressive. It should be
mentioned, however, that the number of stars detected in WFPC2 may
have been improved with less conservative signal-to-noise cuts in
DAOPHOT II, or by using other reduction packages, such as HSTPHOT. 

\epsscale{1.00}
\noindent
\begin{figure*}
\plotone{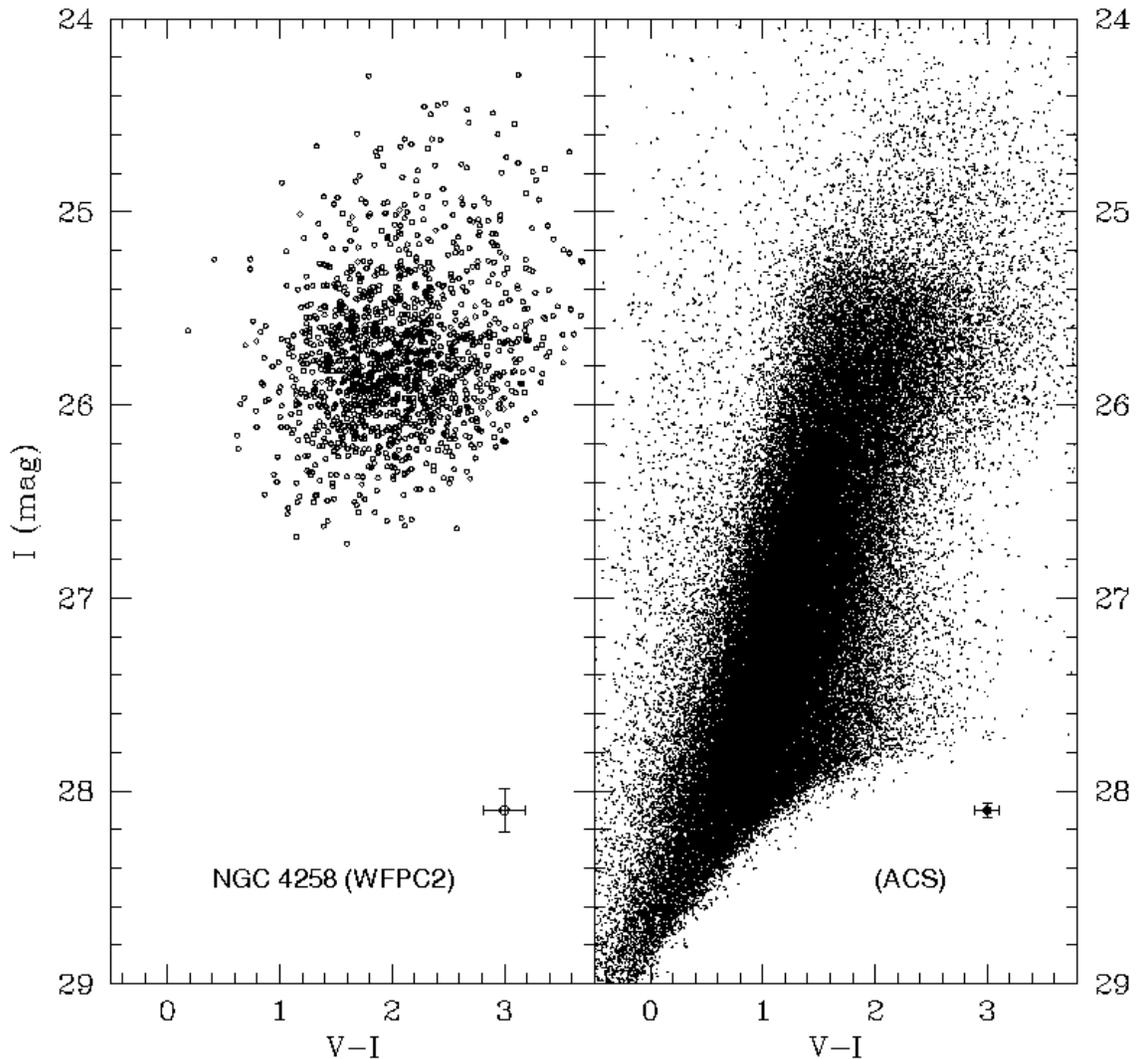}
\caption{Color-magnitude diagrams for NGC~4258 from the
WFPC2 data (left panel) and the ACS data (right panel). The error bar
in the lower right corner of each panel represents the median uncertainty
on the photometry of the stars within $\pm 0.3$ mag of the TRGB.
}
\end{figure*}

As can be seen in these figures, there is a readily apparent
luminosity above which the number density of stars drops off
precipitously. We identify this discontinuity with the TRGB, which is
the result of the core helium flash of red giant stars occurring at
about the same bolometric luminosity for stars of all ages $\gtrsim 2$
Gyr. The I-band magnitude of this discontinuity is known to be only
weakly dependent on metallicity for sufficiently low metallicities
([Fe/H] $\lesssim -0.7$ dex) (Iben \& Renzini 1983; Lee et
al. 1993). At higher metallicities, line-blanketing effects begin to
have a more significant effect, and corrections have been suggested to
take this into account (e.g., Salaris \& Cassisi 1998;
Bellazzini et al. 2001, 2004; however, see below and Madore et
al. 2008 for the latest calibration). As such, measuring the magnitude
of this number-density drop-off, which corresponds to the tip of the
red giant branch, has proven to be a reliable way of finding the
distance modulus of any galaxy that has resolvable stars which are
part of an existing population of old red giant stars.

We now consider the uncertainty in measuring the location of the
TRGB. The most important factors include random photometric errors,
sample size, crowding issues, and contamination/confusion caused by
asymptotic giant branch (AGB) stars (Renzini 1992; Madore \& Freedman
1995).  Lee et al. (1993) first introduced a quantitative method of
measuring the magnitude of the TRGB and its uncertainty. They used a
zero-sum (Sobel) kernel edge-detector [-1, 0, 1] applied to the binned
histograms of the observed luminosity functions. This filter produces
a maximum response at the magnitude where the slope of the luminosity
profile is largest. However, this method is sensitive to random
noise spikes in the luminosity function, and will, of course, produce
slightly different answers depending on the choice of bin size, the
starting-point of the histogram, etc. (Madore \& Freedman 1995; Sakai
et al. 1996). To reduce the impact of noise spikes, Madore \& Freedman
used a modified version of the Lee et al.  edge-detector, with a
weighted Sobel filter that smoothed the data over 2 bins on either
side of the central bin (i.e., [-1, -2, 0, +2, +1]). Sakai et
al. (1996) went a step further, Gaussian-smoothing the luminosity
distribution and applying an edge-detection filter to the continuous
function, thereby avoiding the issues involved in the discrete
binning. M\'{e}ndez et al (2002) modified this method to take into
account the natural power-law distribution of the luminosity function,
using a logarithmic ratio in their edge-detection filter instead of
the first derivative. Additionally, they employed a maximum-likelihood
analysis as an alternative method of estimating the position of the
TRGB, with uncertainties being derived from bootstrap re-sampling of
the data. Mouchine et al. (2005) also used the maximum-likelihood
analysis and bootstrap re-sampling to find TRGB distances to several
galaxies, including NGC~4258. Other authors (e.g., Cioni et al. 2000;
Sarajedini et al. 2002; Frayn \& Gilmore 2003; McConnachie et
al. 2004; Makarov et al. 2006) have additionally modified these and 
similar techniques,
reducing the contribution of random noise spikes in the luminosity
function. Madore et al. (2008) discuss yet another modification to the
basic technique, this time aimed at capturing the metallicity-sensitivity 
of the TRGB itself. A similar methodology is described and applied below.

One problem inherent in all of these methods is that for some
galaxies, the edge-detector can produce multiple peaks (some of which
are larger than the one at the location of the TRGB itself). Choices
were then made by eliminating peaks based on a priori knowledge of the
general location of the TRGB. It is most desirable to remove this
ambiguity, and develop an algorithm that can produce a reliable result
without added intervention.  For this analysis, we employ another
method of measuring the magnitude of the TRGB that is robust against
random noise spikes in the luminosity profile, yet does not rely on
fitting the data to any adopted model.

Uncertainties are introduced in the TRGB distance modulus by the
metallicity dependence of the magnitude of the TRGB (Bellazzini 2001).
This can be seen as a color dependence of the TRGB magnitude that is
apparent in the ACS CMD in Fig. 2 (right panel): the edge associated
with the TRGB is sloped slightly downward, becoming fainter at redder
colors. Past studies have dealt with this issue by fitting a Gaussian
to the color distribution of the TRGB stars, and using the peak of
this distribution to determine the average metallicity of the stars,
which in turn was used to correct the tip magnitude to a fiducial
metallicity/color. We side-step this process by applying a metallicity
correction to each star before running the edge-detection software.

In the following sections we discuss our modified version of a
metallicity-corrected Sobel edge-detector, and use it to determine our TRGB 
distance to NGC~4258.

\subsection{Correcting for Metallicity Dependence}
 
Bellazzini et al. (2001, 2004) re-affirmed that both the absolute
magnitude (M$_I^{TRGB}$) and color of a star at the tip of the red
giant branch are functions of metallicity as given by the following
relations:
\begin{equation}
M_I^{TRGB} = 0.14 [Fe/H]^2 + 0.48 [Fe/H] - 3.629
\end{equation}
\begin{equation}
(V-I) = 0.581 [Fe/H]^2 + 2.472 [Fe/H] + 4.013
\end{equation}

Clearly, these equations can be solved simultaneously to obtain
M$_I^{TRGB}$ as a function of the tip (V--I) color. We have
numerically solved these simultaneous equations for the run of
absolute magnitude with color. Those data points are shown in Fig. 3;
the relation is clearly non-linear, but two rough linear approximations
are shown (note that there is a secondary solution to the resulting
quadratic equation that has been rejected here as it does not fit the data 
or theory well). The dashed line has a slope of 0.20, and is a very good 
approximation to the plotted
points red-ward of (V-I) = 2.0~mag, but it systematically deviates to
brighter magnitudes by up to 0.1~mag at (V-I) = 1.6~mag. However, it 
approximates the theoretical points over the entire color range of interest 
with a residual scatter of only $\pm$0.026~mag. This slope also has
observational support from Rizzi et al. (2007). The solid line shows
a linear solution of slope 0.15 over the entire TRGB color range seen in 
nature, but it does not fit the theoretical data points as well.
The linear approximation with slope = 
$0.20 \pm 0.05$ is used in Madore et al. (2008), as it agrees within the
error with both the observationally determined TRGB slope, and the  
linear fit to the analytical solution (pictured by the circled points
in Fig. 3). For accuracy, we apply the full analytical solution to the
data for this paper, although the linear approximation (like that given in 
Madore et al. (2008)) is a valid alternative. 


\epsscale{1.00}
\noindent
\begin{figure*}
\includegraphics[angle=-90]{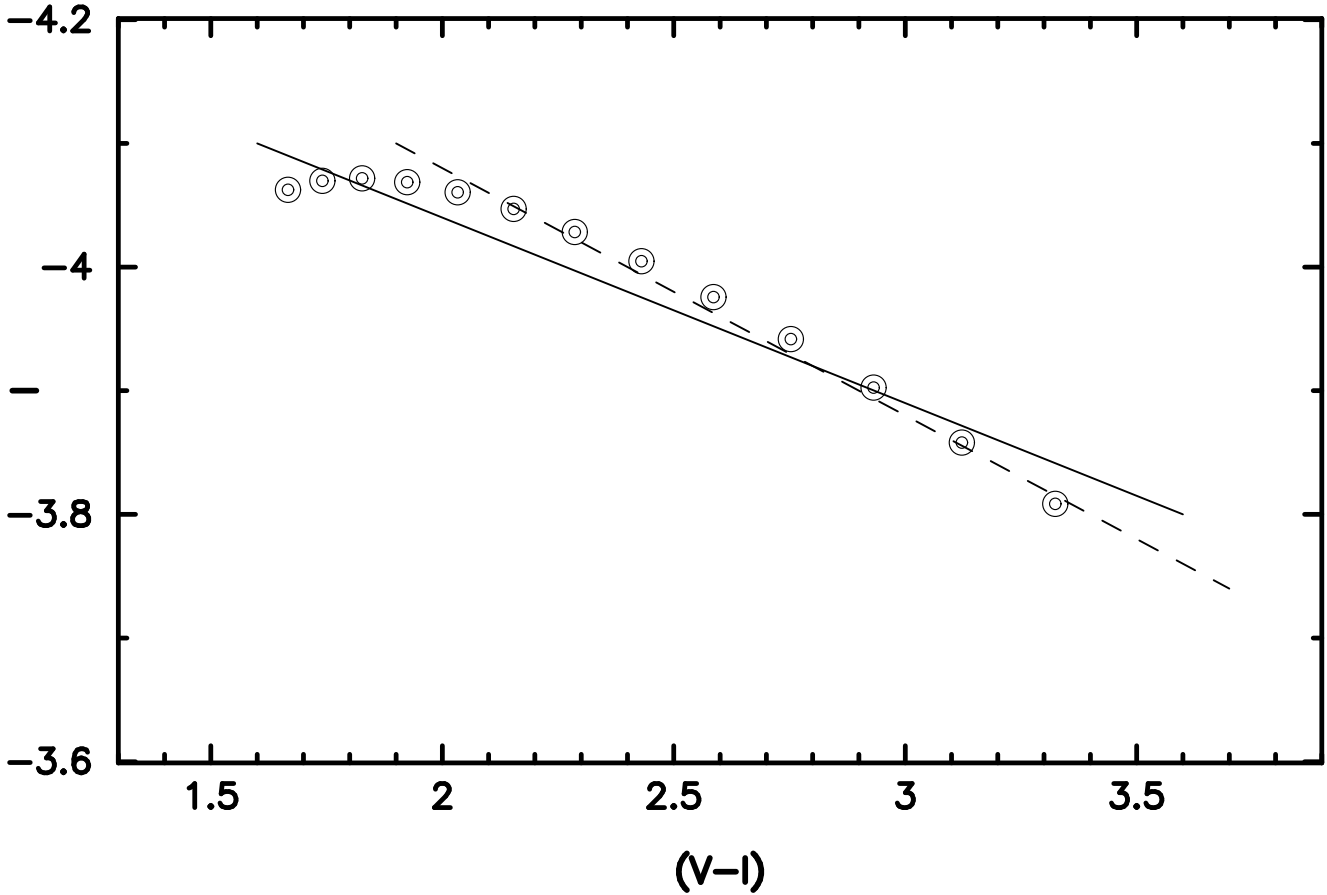}
\figcaption{The theoretical dependence of the absolute I-band
magnitude of a TRGB star on its (V--I) color, as derived from the
relations of M$_I^{TRGB}$ and (V--I) vs. metallicity (Bellazzini et al.
2001, 2004). The solid line shows a rough fit over the color range
(V--I) $\sim$ 1.6--3.6, with a drop of 0.15 mag/mag. The dotted line has a
slope of 0.20 mag/mag which is a closer fit to theory for colors redder
than $\sim 2.0$, and which is more closely supported by the observations of
Rizzi et al. (2007). One is left with
the following choices: (a) a linear approximation of shallow (0.15)
slope representing the (non-linear) trend over the entire color range
seen for TRGB stars, (b) a steeper linear approximation
of slope 0.15 which more closely approximates the theory (and the
restricted observations), but only over a more narrowly-defined and
redder range of color than observed, or (c) applying the
detailed non-linear theoretical correction to the data over the full
color range.  Striving for accuracy, we use method c, removing this color
dependence in our analysis and arbitrarily normalizing the TRGB stars to the
I-band magnitude of a TRGB star with a color of (V--I) = 1.6.
}
\end{figure*}


Armed with an equation describing the magnitude of the TRBG with color
we can now correct the magnitudes of all stars for metallicity before
running the tip detection algorithm down through the color-magnitude
diagram.  With this correction applied, we find consistent results in
the ACS data with color for the I-band apparent magnitude of the TRGB
(I$_{TRGB}$) in sub-samples of stars over the range $1.6 \lesssim$
(V--I) $\lesssim 3.0$ (see Section 4.2 for an explanation of how
I$_{TRGB}$ is measured). As such, we have effectively removed the
metallicity dependence of the I$_{TRGB}$ measurement and implicitly
normalized the resulting I-band magnitudes to that of a TRGB star with
(V--I) = 1.60~mag. This corresponds to [Fe/H] = --1.52, and
M$_I^{TRGB}$ = --4.04 $\pm 0.12$~mag (with the uncertainty on
M$_I^{TRGB}$ determined by Bellazzini et al. 2001, 2004).  As can be
seen in the left panel of Fig. 5, this transformation (by
construction) removes the downward slope of the TRGB edge that is
clearly seen in the right panel of Fig. 2.

This process offers an improvement over previous methods of using the
average color of the TRGB stars to determine the average value of
M$_I^{TRGB}$. It also eliminates the need to apply a red color cut-off
to the TRGB, which has previously been done to avoid the region where
the TRGB magnitude is increasingly depressed by metallicity and/or
contaminated by extended AGB stars. This increases the fraction of
stars usable in the edge-detector, and thus decreases the uncertainty
of the result through higher number statistics.
 
\subsection{Weighted Edge-Detector and Results from the ACS data}

As in M\'{e}ndez et al. (2002), we accommodate the power-law
distribution of the luminosity profile by utilizing a logarithmic
ratio in our edge detector. For the purpose of eliminating the
necessity to fit the data to theoretical models, however, we apply
this directly to the histogram of the I-band luminosity distribution,
and test the results as a function of bin-size. We also normalize the
edge-detector output by the Poisson noise, expected from $\sqrt{N}$
statistics. The basic form of the edge-detector yields an output
filter response at the ith magnitude bin of:
\begin{equation}
\eta_i^o = \sqrt{N} \times (log(m_{i+1}) - log(m_{i-1})),
\end{equation}
\noindent where N is the number of stars in the central {\it i}th
magnitude bin.  The magnitude bin corresponding to the maximum peak of
$\eta_i^o$ gives the magnitude at which the change in luminosity from
one bin to the next is the largest. We identify that maximum response
with the TRGB.

We tested this basic edge-detector by applying it to the luminosity
histogram of the metallicity-corrected NGC~4258 ACS data. To eliminate
contamination from non-RGB stars, we included only stars with colors
corresponding to those of the TRGB, and for which the measured value
of I$_{TRGB}$ from the weighted edge detector (as given by Eq. 5) is
consistent with sub-selections in color ($1.6 <$ (V--I) $< 3.0$).  The
I-band histogram of these data is presented in Fig. 4 (a), using a bin
size of 0.02 mag. Fig. 4 (b) shows the basic edge detector response
($\eta^o$ from Eq. 4) to this histogram. At this resolution the
detector response is noisy, with multiple peaks having comparable
significance. The maximum peak is at I = 25.20 mag, but it is
uncompelling.  This is the case even though we have exquisite number
statistics, with over 10,000 stars in the 1 mag bin below the
TRGB. Past studies with shallower data have typically had number
statistics close to or below the statistical limit at which the TRGB method 
was considered to be reasonably accurate (i.e., $50-100$ stars in the single 
magnitude bin below the TRGB, Madore \& Freedman 1995). Problematic to these
past studies, a recent analysis has found an even stricter statistical
limit of 400-500 stars (Madore et al. 2008). The filter response is even 
more noisy for data
with lower number statistics, in which case taking different
sub-samples of stars or even slightly changing the starting magnitude
and bin size can cause one of the other peaks to be randomly
higher. In some cases it can yield a vastly different answer for the
magnitude of the steepest edge. For instance, we see this effect in
our data if we choose a smaller sub-sample of stars by making a color
cut of $1.6 <$ (V--I) $< 2.0$. There are still more than enough stars to meet the
minimum requirement for this method (i.e., 5,900 stars in the mag bin
below the TRGB), but the highest peak in the edge detector response is
now at 26.36 mag, which corresponds to the second highest peak, previously seen  in
Fig. 4(b). This is a typical problem with Sobel detectors (even those
that fit the data to a smooth curve), and has been alleviated by some
in the past by picking the peak that corresponds to the most likely
location of the TRGB as seen in the CMD. It is desirable to remove
this ambiguity by modifying the edge detector so that a priori
assumptions about the general location of the magnitude of the TRGB
are unnecessary, even in data-sets with lower number statistics.


\epsscale{1.00}
\noindent
\begin{figure*}
\plotone{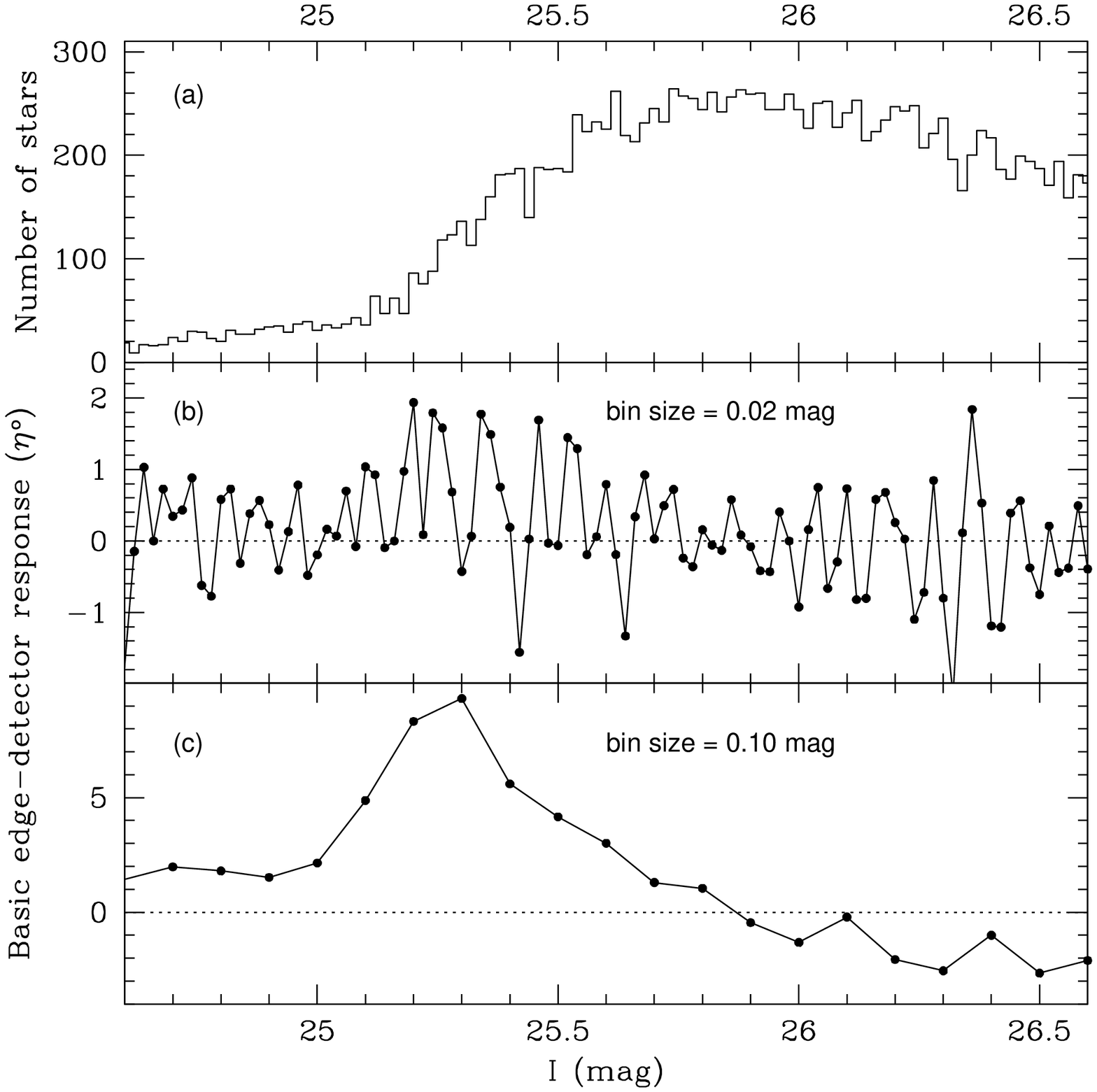}
\caption{{\bf(a)} Luminosity distribution of stars in the
ACS images of NGC~4258 with colors corresponding to those of the TRGB
($1.6 <$ (V--I) $< 3.0$). Bin sizes are 0.02 mag. {\bf(b)} Basic
edge-detector response, as defined by Eq. 4, for the histogram in (a).
This detector response is highly dependent on noise-spikes within the
luminosity profile, and may require a priori knowledge of the location
of the TRGB to determine which peak corresponds to it.
{\bf (c)} Basic edge-detector response for a histogram with bin
sizes of  0.10 mag bins. This detector response is more robust to noise,
yielding a single unambiguous peak near the true location of the TRGB.
However, it is not as precise as a filter applied to smaller magnitude
bin sizes. We convolve the filter response in (b) and (c) into one robust,
yet precise filter in Eq. 5 and Fig. 5.
}
\end{figure*}


If we use wider bins, we effectively smooth the data and thereby
reduce the noise, at the cost of precision of our result and
potentially washing out important structures within the luminosity
profile. We now explore this option. Using bins of size 0.05 mag still
leaves ambiguous spikes, as seen when using the 0.02 mag bins. These
ambiguous cases begin to disappear when we use bins 0.10 mag in size.
As shown in Fig. 4(c), there is now only one statistically significant
peak at $\sim 25.30$ mag, which is roughly near the by-eye location of
the TRGB in the CMD in Fig. 5. Unlike the 0.02 bin-size edge detector,
this result is robust when testing it on smaller sub-samples of
stars. Using bins much larger than 0.10 mag over-smoothes
the data, and the TRGB edge becomes undetectable.

We can achieve the best of both worlds by convolving the high
precision 0.02 mag bin-size filter (which has high precision but is
sensitive to noise) with the 0.10 mag bin-size filter (which is
over-smoothing the data, but is robust to high-frequency noise-spikes
in the luminosity profile). We do this for a histogram with original
bin widths of 0.02 mag, by modifying Eq. 4, such that we take the
logarithmic ratio of the sum of the 5 bins fainter and 5 bins brighter
than the {\it i}th magnitude bin. We then step the filter by 0.02 mag
in turn, producing a filter response, $\eta$, for every 0.02 mag in
the histogram. We again normalize this output by the Poisson noise in
the central three bins in order to properly reduce significance of the
noise-induced fluctuations in the filter response. The new modified
equation for the weighted edge-detector response at the ith magnitude
bin is:
\begin{equation}
\eta_i = \sqrt{\sum_{i-1}^{i+1} N_n} \times (log(\sum_{i+1}^{i+5} m_{n}) 
- log(\sum_{i-5}^{i-1} m_{n})).
\end{equation}

Fig. 5 displays the results of applying this weighted edge-detection
filter to the NGC~4258 ACS data. The left panel of Fig. 5 shows the
CMD with the metallicity correction derived from Eq.'s 2 and 3 applied
to each star. The right panel shows the edge-detector filter response,
$\eta$, from Eq. 5. The shape of this response is a combination of the
large-scale response from the 0.10 mag bin basic filter from Eq. 4
(see Fig. 4 (c)), and the higher precision, small-scale filter
response of the 0.02 mag bin basic filter (Fig. 4 (b)). The result is
an unambiguous maximum peak at I = 25.24 mag, which is marked on
Fig. 5 with a horizontal dashed line.


\epsscale{1.00}
\noindent
\begin{figure*}
\plotone{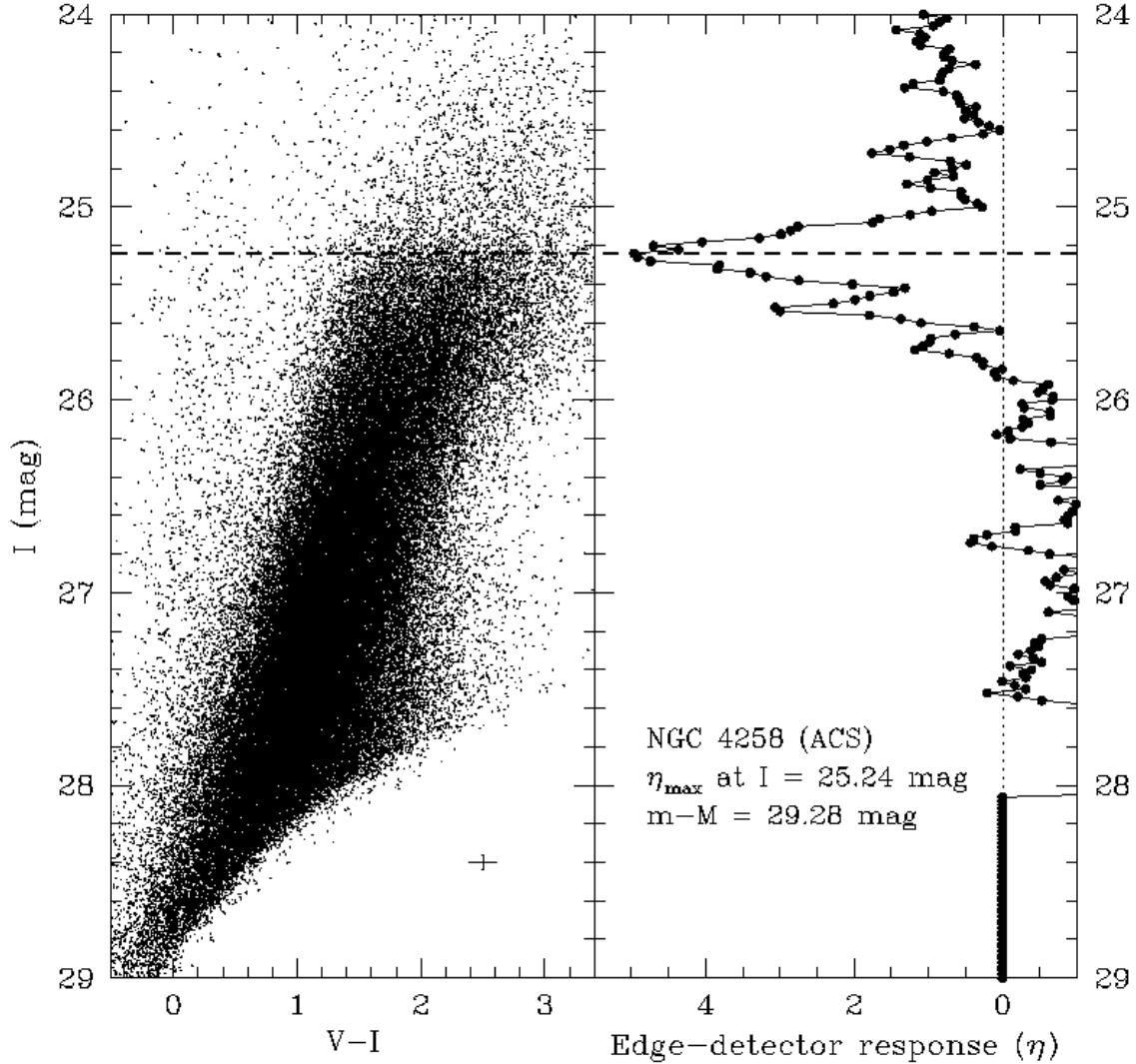}
\caption{{\bf Left:} The CMD from the NGC~4258 ACS data after
applying the TRGB metallicity correction to each star (as derived from
Eq.'s 2 and 3). {\bf Right:} The value of the edge-detector response ($\eta$)
from Eq. 5. The maximum peak of $\eta$ gives our measured apparent
magnitude of the TRGB, which is marked by the dashed horizontal line in
both panels.
}
\end{figure*}


\subsection{Error Analysis for the ACS data}

We determined the statistical uncertainty on the I$_{TRGB}$
measurement by applying our edge-detector algorithm to 63 sub-samples
of stars, selected as a function of color, and with various sample
sizes achieved by ranging the width of the selected color bins from
0.2 to 1.4 mag. Random variations in the result due to statistical
fluctuations in the luminosity profile should be apparent from this
test.  We find a median value for I$_{TRGB}$ of 25.24 mag within these
sub-samples, which is the same as the result from the larger
sample. There are variations in the solution among individual
sub-samples, however, with the largest variation seen in smaller color
bin-size samples, which subsequently have smaller number
statistics. Of the 63 sub-samples, 6 have I$_{TRGB}$ measurements that
are clear outliers, either around I $\sim 25.5$, or $I \sim
24.3$. These occur in relatively less populated bins, and are likely
due to secondary peaks in the luminosity profile of these particular
sub-samples being randomly higher than the peak at the location of the
TRGB. In the case of data-sets with poor stellar number statistics, we
therefore recommend applying a similar test in order to reject any
outlying secondary solutions. Of the remaining 57 sub-samples, the
individual measurements of I$_{TRGB}$ vary by as much as $\pm 0.10$
mag from the median in the 0.2 mag color bin size samples, $\pm 0.06$
mag in the 0.3--0.4 mag bin size samples, and $\pm 0.04$ mag in the
0.5--1.4 mag bin size samples. For our measurement of I$_{TRGB}$ from
the largest sample ($1.6 <$ (V--I) $< 3.0$), we adopt an uncertainty
corresponding to the range of values achieved in the larger
sub-samples, giving I$_{TRGB}$ = $25.24 \pm 0.04$ mag.

Systematic errors also contribute to the uncertainty of I$_{TRGB}$.
One such possible error source is stellar crowding, which can lead to
systematically brighter measurements for the TRGB (Madore \& Freedman
(1995)). To test the effects of crowding on our results, we applied
our weighted edge detector to three separate sections of both chips in
the ACS images, each with a different density of stars based on its
proximity to the disk of the galaxy. The detected edge in each of
these image sections varied within the statistical uncertainty of the
detector ($\pm 0.04$ mag), and did not show any trend with increasing
stellar density. Therefore, we conclude that crowding does not have a
significant systematic effect in this ACS field.

\subsection{Results from the WFPC2 Data}

As a consistency check, we also applied our weighted edge-detector to
the WFPC2 images of NGC~4258, providing a direct comparison of our
measured TRGB apparent magnitude to that of Mouhcine et al. (2005)
from the same data set. As evident in Fig. 1, the WFPC2 observations
spatially overlap with the ACS observations, and thus sample largely
the same population of halo stars. As can be seen in Fig. 2, however,
there are far fewer stars detected in the WFPC2 images than the ACS
images (mostly due to the superior sensitivity of the ACS detector).
As such, the WFPC2 data will provide us with an indication of the
applicability of our edge-detector to data with lower number
statistics.  Even so, the WFPC2 images still exceed the minimum
requirements for a reliable measurement of the TRGB magnitude (Madore
\& Freedman 1995, Madore et al. 2008), with slightly less than 1000 stars 
in the 1 mag-width bin brighter than the TRGB (even after the application of
our color and $\chi$ value limits, as described below and in Section
3, respectively).

The metallicity-corrected CMD of the WFPC2 data is presented in the
left panel of Fig. 6. As with the ACS data, we test the edge detector
on 63 sub-samples of these stars as selected by color, and with
various sample sizes within 0.2--1.4 mag bins in color. Because the
WFPC2 data is shallower in V than the ACS data, we find that
sub-samples including stars with (V--I) $\gtrsim 2.8$ mag lead to
detected edges that are significantly fainter (by $\gtrsim 0.2$ mag)
than the values obtained from bluer sub-samples of stars. As such, we
apply slightly bluer color cuts to this data of $1.4 <$ (V--I) $<
2.8$.  The edge-detector response ($\eta$) from Eq. 5 for these data
is shown in the right panel of Fig. 6. Although the detector response
is noisier than that of the ACS data (see Fig. 5), the maximum peak is
near the by-eye location of the TRGB, at I$_{TRGB}$ = 25.20 $\pm 0.06$
mag, which is consistent with the median value of I$_{TRGB}$ in the
color-selected sub-samples. As with the ACS data, the uncertainty is
obtained from the range of values of the detected edges in each of the
sub-samples after rejecting extreme outliers.  I$_{TRGB}$ = $25.20 \pm
0.06$ mag, as measured from the WFPC2 data, is 0.04 mag brighter than,
but fully consistent with, I$_{TRGB}$ = $25.24 \pm 0.04$ mag, as
measured from the ACS data. Madore \& Freedman (1995) find that photometric
errors and crowding can cause the discontinuity attributed to the TRGB
to appear brighter than it really is.  The WFPC2 data do have larger
photometric errors than the ACS data, although crowding is not as much
of an issue in the shallower WFPC2 images.


\epsscale{1.00}
\noindent
\begin{figure*}
\plotone{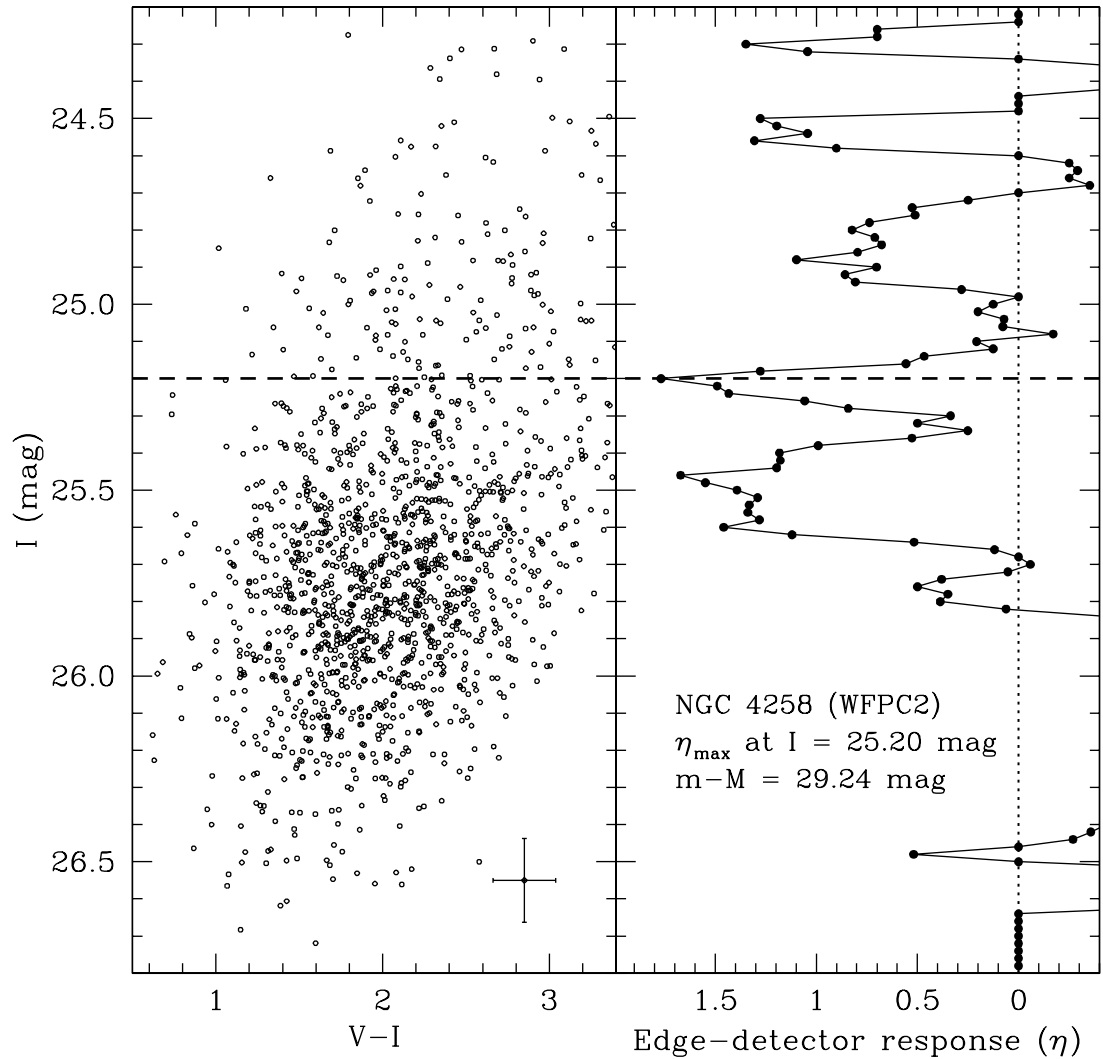}
\caption{{\bf Left:} The CMD from the NGC 4258 WFPC2 data after
applying the TRGB metallicity correction to each star (as derived from
Eq.'s 2 and 3). {\bf Right:} The value of the edge-detector response ($\eta$)
from Eq. 5. The maximum peak of $\eta$ gives our measured apparent
magnitude of the TRGB, which is marked by the dashed horizontal line
in both panels. This agrees with the result from the ACS data within
the uncertainties.
}
\end{figure*}


\section{The TRGB Distance Modulus; Comparison to Previous Results}

As explained in Section 4.1, the absolute magnitude of our
metallicity-corrected TRGB stars is taken to be M$_I^{TRGB}$ = --4.04
$\pm 0.12$ mag.  This leads to a distance modulus of (m--M)$_{ACS} =
29.28 \pm 0.04$ (random) $\pm 0.12$ (systematic) mag, and
(m--M)$_{WFPC2} = 29.24 \pm 0.06$ (random) $\pm 0.12$ (systematic)
mag. The ACS result is in exact agreement with the distance modulus of
29.28 $\pm 0.09$ mag that was obtained through the Keplerian motion of
nuclear water masers orbiting the central black hole (Herrnstein et
al. 1999). The distance obtained through the orbital motions of these
masers relies on simple, well-understood basic principles, and as such
this distance determination for NGC~4258 carries high
weight. Unfortunately, with only one example of this method to work
with it is extremely hard to externally assess the systematics of this
method. Hopefully more examples will be forthcoming.

As another consistency check, we compare our TRGB results to that of
Mouhcine et al. (2005), who independently found the TRGB distance
modulus from the same WFPC2 field we used for this analysis.  Mouhcine
et al. determined the apparent I-band magnitude of the TRGB by
Gaussian-smoothing the luminosity function of all stars in the WFPC2
field with (V--I) $< 2$, and applying both a continuous function Sobel
edge-detector, and maximum likelihood analysis. They then measured the
average metallicity of their color-selected stars by fitting a
Gaussian to the color distribution and applying the result to the
metallicity-color relation from Lee et al. (1993). This average
metallicity was then used in the distance modulus vs. I$_{TRGB}$
equations in Lee et al. to find (m--M)$_{\circ}$ = $29.32 \pm 0.09$ (random)
$\pm 0.15$ (systematic) mag. This is 0.04 mag fainter than our result
from the ACS data, and 0.08 mag fainter than our result from the WFPC2
data. However, despite different methods used in the edge-detector
algorithm and treatment of the metallicity dependence of the TRGB
magnitude, all of these results agree within the
uncertainties. This apparent confirmation, on the other-hand, is not all
that robust.

There is less agreement with Macri et al. (2006), who used their outer 
disk CMD data in an attempt to measure a TRGB distance.  
They noted a detection at I = 24.42
$\pm$ 0.02~mag, which led them to a distance modulus of 29.41 $\pm$
0.04~mag.  Examination of their published CMD shows that they were
working very close to their detection limit, and it is not clear how
pure the Population~II component would be in a region chosen for
Population~I Cepheid discovery.  However, there is independent support
for a large distance modulus. 

Even more recently Rizzi et al. (2007) have published a pre-analysis 
of our data on NGC~4258 using their own
TRGB detection methods and metallicity calibration. They claim a tip
detection at 25.49~ $\pm$ 0.05~mag, corresponding to a 
metallicity-corrected distance modulus of 29.42 $\pm$ 0.06~mag. These moduli
respectively are 7 and 3 sigma away from our value. Rizzi et al. (2007) note
that their value puts them in very good agreement with the Macri et
al. (2006) Cepheid distance modulus, but it places them 2-sigma away
from the maser distance. To address this discrepancy, Macri et al. 
re-calculated their TRGB tip magnitude using the same metallicity correction,
color limits, and reddening correction that were applied in this paper 
(Tully, B., private communication). They found a revised I-band tip
magnitude of 25.26 mag, which is in excellent agreement with our value 
of $25.24 \pm 0.04$ mag. They also find a difference of 0.09 mag in
the results of their own analysis when adjusting for metallicity before
vs. after running their tip finder. This indicates that the 0.14 mag 
difference between the published Rizzi et al. (2007) distance modulus 
and our own is likely not due to the tip detection method itself, 
but to systematic differences in the application of the metallicity 
correction. This results in a systematic uncertainty associated with
the calibration zero point, which is encompassed by our quoted 
systematic uncertainty of $ \pm 0.12$ mag. 

Prompted by these differences, we now take an independent look at the 
Cepheid data.

\section{The Cepheid Data for NGC~4258}


\noindent
\begin{figure*}
\includegraphics[width=14cm, angle=-90]{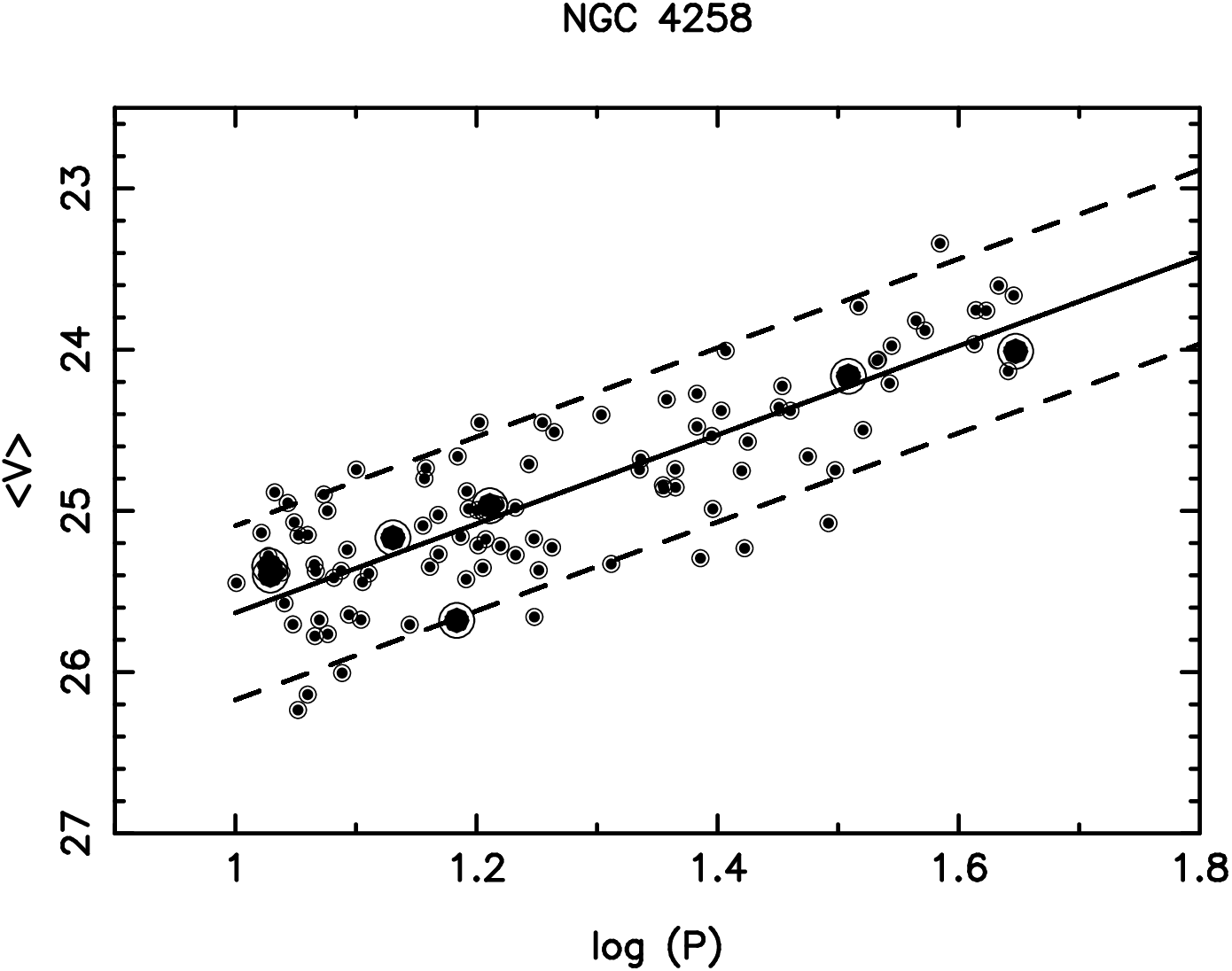}
\figcaption{The V-band PL relation for all long-period (P $> 10$ d)
Cepheids
in NGC~4258 using data from Macri et al. (2006). Small symbols are for
Cepheids in the inner field; large symbols track the Cepheids in the
outer field. The solid line represents the expected trend from
calibrations, with the dotted lines marking the 2-$\sigma$ variation from
this trend (Madore \& Freedman 1991). Note the significantly smaller
scatter of the larger symbols with respect to the expected range.
}
\end{figure*}


An et al. (2007) applied redenning and metallicity corrections     
to Cepheids in NGC~4258, and found a distance modulus of     
m-M = $29.28 \pm 0.10$ (random), $\pm 0.16$ (systematic). While this agrees 
exactly with our value, measurements from other authors are all not in as
close agreement. For example, Macri et. al (2006) present VI data for 
281 Cepheids in two radially separated fields in the disk of 
NGC~4258. From these data they derive
a distance-modulus difference of 10.88 $\pm$ 0.04~mag between NGC~4258
and the LMC. Scaled to the value of the LMC distance modulus
(18.50~mag) adopted by the HST Key Project (Freedman et al. 2001) this
corresponds to a distance modulus of 29.38~mag for NGC~4258. How
robust is this number?


\noindent
\begin{figure*}
\includegraphics[width=14cm, angle=-90]{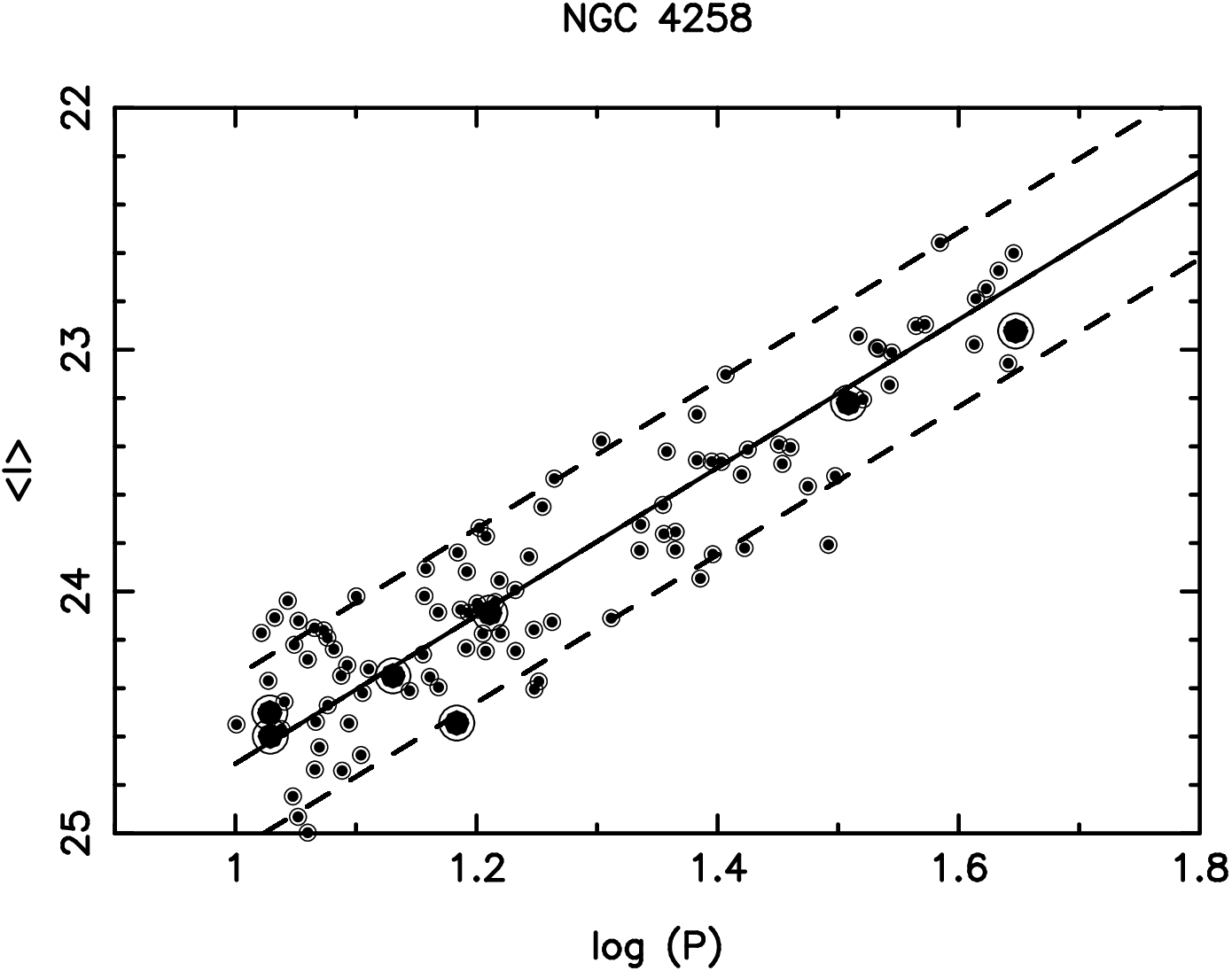}
\figcaption{The I-band PL relation for all long-period Cepheids
in NGC~4258 using data taken from Macri et al. (2006). Small symbols
are for Cepheids in the inner field; large symbols track the Cepheids
in the outer field. The solid line represents the expected trend from
calibrations, with the dotted lines marking the 2-$\sigma$ variation from
this trend (Madore \& Freedman 1991). Note the significantly smaller
scatter of the larger symbols with respect to the expected range.
}
\end{figure*}


So as to be totally consistent with the Key Project zero points and
methodology we restrict ourselves to Cepheids in NGC~4258 that have
periods in excess of 10 days. This has the added advantage of using
stars with the highest signal to noise, and implicitly avoiding other
issues concerning the putative change in slope of the PL
(period-luminosity) relation below 10 days, and the possible contamination 
of the data set by over-tone pulsators that are also only found below 
10 days in period.

Using those restricted samples of Cepheids in each of the two radial
fields, and the method described in Freedman et al. (1994), we find the 
following:

Outer Field: $\mu_V = 29.80 \pm 0.10 $, $\mu_I = 29.66 \pm 0.06 $

Inner Field: $\mu_V = 29.84 \pm 0.04 $, $\mu_I = 29.60 \pm 0.03 $

The very first thing to notice is that the apparent moduli for
these two distinct fields in both bands are statistically the same to
within one sigma of each other. However, there are slight differences,
and these in turn give rise to systematic differences in the formally
derived reddenings for each of the two fields: the Inner Field
Cepheids have a mean reddening of $E(V-I) = 0.24$~mag while the Outer
Field Cepheids have a calculated mean reddening of $E(V-I) =
0.14$~mag. These differences in reddening get further multiplied up (by
the ratio of total-to-selective absorption) yielding systematic
differences in the extinction: $A_V(inner) = 0.58$~mag and
$A_V(outer) = 0.34$~mag. Ultimately the two extinction-corrected
(true) moduli become:

Outer: $\mu_{\circ} = 29.45 \pm 0.08 $

Inner: $\mu_{\circ} = 29.26 \pm 0.03 $

\noindent
This difference represents a 2-sigma significance of separation
amounting to about 0.2~mag. The presumptive interpretation is that
this difference is due to metallicity differences between the inner
and outer field Cepheids.  There are, however, other possible
interpretations of the data. The first is that the reddening law may
be different between the two fields. The second is that small number
statistics have generated the difference. That is, for the small
sample of stars in the outer field we just got (un)lucky. 
We offer up evidence in support of that possibility
below.

First we set aside the possibility of systematically changing the
reddening law for a different paper. The implications of changing the
reddening law from galaxy to galaxy, or from place to place within a
given galaxy are too far-reaching for us to give it the necessary
attention in this paper. Suffice it to say that if the canonical value of 
$R_{VI} \simeq 2.4$ is adopted for the inner field, the two moduli can 
be made to agree if $R_{VI} \simeq 4.0$ in the outer field.
The second possibility is easy to consider,
and it has few consequences beyond this particular application. It is
also motivated by our earlier observation that the apparent moduli in
the two fields were statistically identical but systematically
diverged as reddening corrections were derived and applied.

To shed some light on the possibility that there may be no real
differences in the two Cepheid populations, we now merge the two data
sets and invite the reader to inspect and consider the results. In
Figures 7 and 8 we show the apparent V and I-band PL relations for the
Cepheids in NGC~4258; the large filled circles are the Cepheids in the
outer field, and the more abundant smaller filled circles are Cepheids in
the inner field. Had there been no differences in the symbols the
inner and outer-field Cepheids would be inseparable.  That is, the
outer-field Cepheids would fall entirely within the known bounds of the
fiducial PL relation instability strip; but they do not fill the
strip.  This latter point is important.  Because the outer
Cepheids do not fully sample the strip, their mean is suspect and may
well be biased.  The width of the instability strip that these
particular Cepheids delineate is, in fact, about a factor of two
smaller than the known width, whether that is measured in the I-band,
the V-band or, more importantly, in the reddening-free W-PL
relation (Madore 1982). The latter relation is shown in Figure 9, where the
outer-field Cepheid data points are so tightly co-linear that they
have a measured width that is about a factor of four smaller than the
intrinsic width. 


\noindent
\begin{figure*}
\includegraphics[width=14cm, angle=-90]{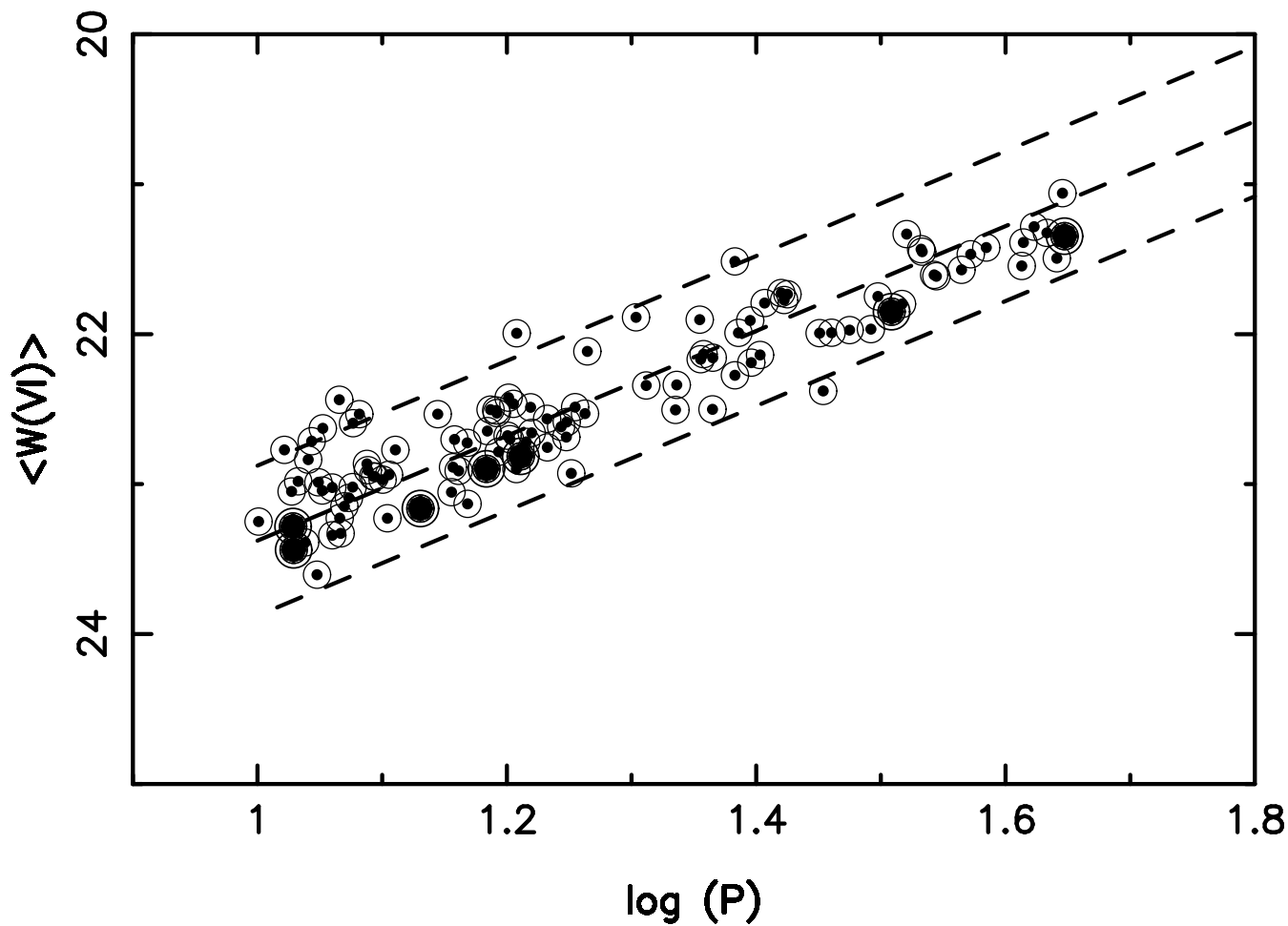}
\figcaption[f9.ps]{The reddening-free W PL relation for all
long-period Cepheids in NGC~4258 using data taken from Macri et
al. (2006). As in the previous two figures the small symbols are for
Cepheids in the inner field. Large symbols track the Cepheids in the
outer field. The solid line represents the expected trend from
calibrations, with the dotted lines marking the 2-$\sigma$ variation from
this trend (Madore \& Freedman 1991). Once again, there is significantly
smaller scatter (this time by nearly a factor of four) of the larger
symbols with respect to the expected range. In this case none of
the scatter can be due to differential reddening.
}
\end{figure*}


We conclude that the outer-field Cepheids are likely to be a biased
sub-set of Cepheids, in that they fail the minimalist requirement of
sampling the entire width of the instability strip before they can be
considered to be a fair sample for either absolute or comparative
purposes.

Using the combined data set of 113 long-period Cepheids, irrespective
of their position in the galaxy, we derive the following apparent
moduli:

 $\mu_V = 29.79 \pm 0.03 $, $\mu_I = 29.58 \pm 0.02 $

\noindent
With a derived mean reddening of $E(V-I) = 0.21 \pm 0.02$~mag, this leads 
to a true distance modulus of

 $\mu_{\circ} = 29.28 \pm 0.02$

\noindent 
This is the true distance modulus to NGC~4258 (corresponding to a metric
distance of 7.18~Mpc) that we believe best reflects the critically
combined Cepheid data. It also coincidentally agrees remarkably well,
indeed exactly, with the independently determined maser distance
modulus of 29.28 $\pm$ 0.09~mag. Moreover it also agrees with our TRGB
distance modulus of 29.28 $\pm$ 0.04~mag. 

Knowing the improbability of these alignments, and given the importance
of these comparisons combined with the published divergence of
solutions, this is probably not to be the last word on
the subject.  However, it is an interesting convergence
of three important distance measurements to a critically important
galaxy.

These results show that the TRGB determination
method described here is reasonably accurate in determining the distance
to this single galaxy. We have also found it to be applicable to several
other, more distant and fainter galaxies, which we will address
in the subsequent papers of this series.

\acknowledgments

This research has made use of the NASA/IPAC Extragalactic Database
(NED) which is operated by the Jet Propulsion Laboratory, California
Institute of Technology, under contract with the National Aeronautics
and Space Administration; and also made use of NASA's Astrophysics
Data System. We thank the referee, Brent Tully, for his constructive
comments.



\end{document}